\documentclass[pra,amsfonts,amssymb,twocolumn,showpacs]{revtex4}
\usepackage{mathrsfs}
\usepackage{graphicx}

\begin{document}

\title{Coherent destruction of tunneling in a lattice array under selective in-phase modulations}

\author{Xiaobing Luo$^{1,2}$}
\altaffiliation{Electronic address: lxbment@yahoo.com.cn}

\author{Jiahao Huang$^{2}$}

\author{Chaohong Lee$^{2}$}
\altaffiliation{Electronic address: chleecn@gmail.com}

\affiliation{$^{1}$Department of Physics, Jinggangshan University, Ji'an 343009, China}

\affiliation{$^{2}$State Key Laboratory of Optoelectronic Materials and Technologies, School of Physics and Engineering, Sun Yat-Sen University, Guangzhou 510275, China}

\date{\today}

\begin{abstract}

We explore the coherent destruction of tunneling (CDT) in a lattice array under selective in-phase harmonic modulations, in which some selected lattice sites are driven by in-phase harmonic oscillating fields and other lattice sites are undriven. Due to the occurrence of CDT, if the driving amplitude $A$ and the driving frequency $\omega$ are tuned to satisfy the zeroth-order Bessel function $J_0(A/\omega)=0$, the driven lattice sites are approximately decoupled with the undriven lattice sites. The CDT even takes place in lattice systems with high-order couplings between non-nearest lattice sites. By using the CDT induced by selective in-phase harmonic modulations, we propose a scheme for realizing directed transport of a single particle. It is possible to observe the CDT in the engineered optical waveguide array, which provides a new opportunity for controlling light propagation and designing switch-like couplers.

\end{abstract}

\maketitle

\section{INTRODUCTION}

Controlling quantum states via external driving fields is a very popular approach for quantum manipulation, which is a basic tool for implementing quantum technology. The external driving fields have been used to manipulate various systems including spins, quantum dots, cold atoms and Josephson junctions etc~\cite{Grifoni,Kohler}. The coherent destruction of tunneling (CDT)~\cite{Grossmann} and dynamical localization (DL)~\cite{Dunlap} are two seminal results in this field. Moreover, some important generalizations of CDT, such as selective coherent destruction of tunneling (SCDT)~\cite{Villas}, non-degenerate CDT~\cite{Stockburger}, nonlinear CDT~\cite{Luo}, and many-body CDT~\cite{Gong}, have also been explored. Single-particle CDT and DL~\cite{Kierig,Lignier}, which have been observed experimentally in optical traps, attract considerable attention due to their potential applications in quantum motor~\cite{Salger,Hai} and quantum information processing~\cite{Romero}. It has also demonstrated a directed transport in a driven bipartite lattice based on the idea of CDT~\cite{Creffield}.

Engineered optical devices offer a new platform for studying quantum dynamical control~\cite{Longhi}. In contrast with a quantum system, an optical device provides the unique advantage of a direct mapping of the temporal evolution of the quantum system into spatial propagation of light waves in engineered lattices. DL in waveguides and CDT in directional couplers are visualized either by periodically bending waveguides~\cite{Longhi2,Valle}, or by an out-of-phase harmonic modulation of refractive index~\cite{Szameit}. These works focus on considering photonic structures with out-of-phase modulations of refractive index of adjacent waveguides~\cite{Szameit}. For a single-band quantum lattice system of periodic modulations on left part of the lattice~\cite{Gong-PRA-2007}, it has demonstrated the occurrence of dissipationless directed transport. However, the question of whether quantum tunneling can be suppressed or even inhibited via selective in-phase modulations is still not clear. Here, the selective in-phase modulation means that the refractive indices for the waveguides oscillate in-phase in selected waveguides, while they are propagation invariant for all other waveguides. Within the context of a driven optical lattice, the in-phase modulation is termed when the selected lattice sites are shaken up and down in-phase while all other lattice sites remain static. In addition, a major limitation of previous schemes for quantum dynamical control is that it is only suitable for systems under nearest-neighbor tight-bing approximation (NNTB), but fails to explain lattices with long-range interaction.

In this article, we analyze how CDT is induced by selective in-phase harmonic modulations and discuss some possible applications of CDT in quantum motor and switch-like coupler. Due to CDT occurs only between the modulated lattice site and the unmodulated ones, one can get the precise location of a particle in the lattice arrays. Under the selective in-phase harmonic modulations, the coupling between a driven lattice site and an undriven lattice site is rescaled by a factor of the zeroth-order Bessel function, while the coupling among the driven lattice sites or the undriven lattice sites keeps unchanged. By tuning the driving frequency and the driving amplitude, the effective tunneling between a driven lattice site and an undriven lattice site can be completely suppressed and therefore the lattice array divides into some sets of disconnected lattice. By using CDT, it is possible to selectively move the quantum particles and design a new type of optical switch-like devices.

The article structure is as follows. In the next section, we give the Hamiltonian for the quantum lattice systems studied in this work. In section III, we explore how CDT is induced by the selective in-phase harmonic modulations. We obtain some analytical results which are confirmed by our numerical simulation. In section IV, we discuss some possible applications of CDT in quantum motor and switch-like coupler. In the last section, we briefly conclude our results.

\section{Hamiltonian}

We consider a single particle in an array of lattice sites, which is described by a single-band tight-binding model. In addition to the tunneling between nearest-neighboring sites, we take account into the high-order coupling between non-nearest-neighboring sites. The Hamiltonian reads as
\begin{eqnarray}
H=&&\sum_j \Omega(\hat{a}^{\dag}_{j+1}\hat{a}_{j}+h.c.)+\sum_j
v(\hat{a}^{\dag}_{j+2}\hat{a}_{j}+h.c.)\nonumber\\ &&+\sum_j
\varepsilon_j(t)\hat{a}^{\dag}_{j}\hat{a}_{j},\label{eq:H}
\end{eqnarray}
where $\Omega$ and $v$ are the tunneling amplitudes connecting nearest-neighboring sites and next-nearest-neighboring sites, respectively.  The $\hat{a}^{\dag}_j$ ($\hat{a}_j$) are the particle creation (annihilation) operators in the $j$-th site.

The in-phase harmonic oscillating fields are applied to modulate the on-site energy $\varepsilon_j(t)$. We consider a type of selective in-phase harmonic modulations of $\varepsilon_j(t) = 0$ for the undriven lattice sites and $\varepsilon_j(t) = A\sin(\omega t)$ for the driven lattice sites. Here, $A$ is the driving strength and $\omega$ is the driving frequency. The case of $\varepsilon_i(t)=\varepsilon_{i+1}(t)=A\sin(\omega t)$ and $\varepsilon_j(t)=0$ for $j\neq i, i+1$ is a typical example of selective in-phase harmonic modulation, where the $i$-th and $(i+1)$-th sites are modulated in phase while all other sites are unmodulated.

\section{Coherent destruction of tunneling induced by selective in-phase modulations}

Taking the Wannier state $|j\rangle=\hat{a}^{\dag}_j|0\rangle$ localized in the $j$th site as basis, we expand the quantum state $|\Psi(t)\rangle$ of  system (\ref{eq:H}) in the form
\begin{eqnarray}
|\Psi(t)\rangle=\sum_{j}a_j(t)|j\rangle, \label{eq:s}
\end{eqnarray}
where $a_j(t)$ represents the occupation probability amplitudes in the site $j$, with the normalization condition  $\sum_j |a_j|^2=1$.

From the Scr\"{o}dinger equation, $i\partial_t|\Psi(t)\rangle=H|\Psi(t)\rangle$, the evolution equation for the probability amplitudes $a_j(t)$ reads
\begin{eqnarray}
i\frac{d
a_j}{dt}=\Omega(a_{j-1}+a_{j+1})+v(a_{j-2}+a_{j+2})+\varepsilon_j(t)a_j.
\label{eq:c}
\end{eqnarray}

The transformation $a_j=b_j\exp[-i\int\varepsilon_j(t)dt]$ yields
\begin{eqnarray}
i\frac{d
b_j}{dt}=&&\Omega b_{j-1}\exp[-i\int(\varepsilon_{j-1}-\varepsilon_{j})dt]\nonumber\\
&&+\Omega
b_{j+1}\exp[-i\int(\varepsilon_{j+1}-\varepsilon_{j})dt]\nonumber\\
&&+v
b_{j+2}\exp[-i\int(\varepsilon_{j+2}-\varepsilon_{j})dt]\nonumber\\
&&+v b_{j-2}\exp[-i\int(\varepsilon_{j-2}-\varepsilon_{j})dt].
\label{eq:ctrans}
\end{eqnarray}

For a $N$-site system under in-phase modulation, the values of $\varepsilon_j(t)$ $(j=1,.....,N)$ are only either $\varepsilon_j(t)=0$ or $\varepsilon_j(t)=A\sin(\omega t)$, so the values of  $ \varepsilon_j(t)-\varepsilon_{j'}(t)$ for $j\neq j'$ should be either $0$ or $\pm A\sin(\omega t)$.  The former case of $\varepsilon_j(t)-\varepsilon_{j'}(t)=0$ indicates that the $j$-th and$j'$-th sites are simultaneously modulated in-phase or unmodulated, and the latter case of $ \varepsilon_j(t)-\varepsilon_{j'}(t)=\pm A\sin(\omega t)$ means that one site is modulated while the other is unmodulated. For the case of $\varepsilon_j(t)-\varepsilon_{j'}(t)=0$, one finds that the effective tunneling parameters between the $j$th and $j'$th sites remain unchanged. For the case of $\varepsilon_j(t)-\varepsilon_{j'}(t)=\pm A\sin(\omega t)$, using the expansion $\exp[\pm iA\cos(\omega t)/\omega]=\sum_k(\pm i)^k J_k(A/\omega)\exp(\pm ik \omega t)$ in terms of Bessel functions and neglecting all orders except $k=0$ in the high frequency region, one finds that the effective tunneling parameters between the $j$-th and $j'$-th sites are renormalized by a factor of $J_0(A/\omega)$.

In our analysis, we focus on the dynamics for three sites. However, similar behaviors may appear in other lattice systems as well. For a three-site system, the coupled equation (\ref{eq:c}) reads as
\begin{eqnarray}
i\frac{d a_1}{dt}=\Omega a_2+va_3+\varepsilon_1(t)a_1,
\nonumber\\
i\frac{d a_2}{dt}=\Omega a_1+\Omega a_3+\varepsilon_2(t)a_2,\nonumber\\
i\frac{d a_3}{dt}=\Omega a_2+v a_1+\varepsilon_3(t)a_3. \label{eq:c3}
\end{eqnarray}

We consider two typical in-phase modulations for the above three-site system: (a) $\varepsilon_1(t)=A\sin(\omega t)$, $\varepsilon_2(t)=\varepsilon_3(t)=0$; and (b) $\varepsilon_1(t)=0$, $\varepsilon_2(t)=\varepsilon_3(t)=A\sin(\omega t)$. In the high-frequency limit, taking time average of high frequency terms, we find the same effective equation of motion for both modulations,
\begin{eqnarray}
i\frac{d b_1}{dt}&=&\Omega J_0\left(A/\omega\right)b_2+vJ_0\left(A/\omega\right)b_3, \nonumber\\
i\frac{d b_2}{dt}&=&\Omega J_0\left(A/\omega\right) b_1+\Omega b_3, \nonumber\\
i\frac{d b_3}{dt}&=&\Omega b_2+vJ_0\left(A/\omega\right)b_1. \label{eq:ceff}
\end{eqnarray}
According to the Floquet theorem, the approximate Floquet solutions of Eq.(\ref{eq:c3}) can be constructed as $a_j=\exp[-i\int\varepsilon_j(t)dt]b_j(t)$ $(j=1,2,3)$ with $(b_1(t),b_2(t),b_3(t))^T=(B_1,B_2,B_3)^T\exp(-iEt)$ being the stationary solution for Eq.~(\ref{eq:ceff}).

\begin{figure}[htp]
\center
\includegraphics[width=1.0\columnwidth]{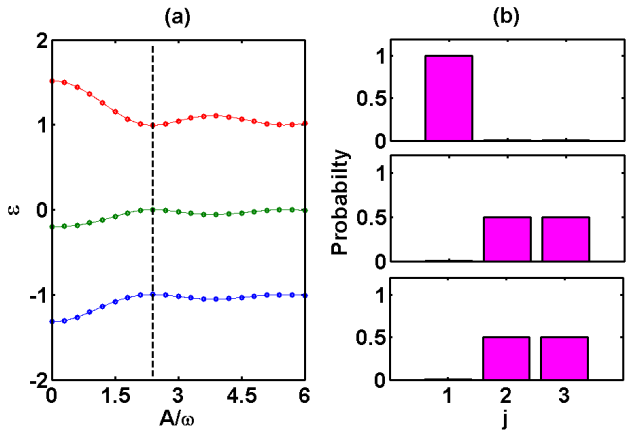}
\caption{(color online) (a) Quasienergies versus $A/\omega$.
The vertical dashed line labels the position of $J_0(A/\omega)=0$. The solid lines and circles correspond to type-(a) and type-(b) modulations, respectively. (b) The probability distribution of the Floquet states for the system of $J_0(A/\omega)=0$ labeled by the vertical dashed line in (a). The other parameters are chosen as $\Omega=1$, $v=0.2$ and $\omega=10$.} \label{fig1}
\end{figure}

If $J_0(A/\omega )=0$, $b_1(t)$ does not couple with $b_2(t)$ and $b_3(t)$. This means that the population in state $\left|1\right\rangle$ does not change with time. However, $b_2(t)$ and $b_3(t)$ are still coupled and so that the populations in states $\left|2\right\rangle$ and $\left|3\right\rangle$ may exchange. There are several roots for $J_0(A/\omega )=0$ such as $A/\omega \simeq 2.4$ and $5.5$. For a system of type-(a) modulation and $J_0(A/\omega)=0$, the approximate Floquet solutions for Eq. (\ref{eq:c3}) read as
\begin{eqnarray}
|u_1(t)\rangle&=&\exp[iA\cos(\omega t)/\omega]|1\rangle+0|2\rangle+0|3\rangle, \nonumber\\
|u_2(t)\rangle&=&[0|1\rangle+\frac{1}{\sqrt{2}}|2\rangle+\frac{1}{\sqrt{2}}|3\rangle]\exp(-i\Omega t),\nonumber\\
|u_3(t)\rangle&=&[0|1\rangle+\frac{1}{\sqrt{2}}|2\rangle-\frac{1}{\sqrt{2}}|3\rangle]\exp(i\Omega
t). \label{eq:fl}
\end{eqnarray}
Clearly, the Floquet state $|u_1(t)\rangle$ is localized in the site 1, and $|u_{2,3}(t)\rangle$ are localized in the sites 2 and 3 with equal probability. In Fig. 1, we show the dependence of the quasienergy on the ratio $A/\omega$. The quasienergy spectra are identical for type-(a) (solid lines) and type-(b) (circles) modulations. The vertical dashed line corresponds to $J_0(A/\omega )=0$. In Fig. \ref{fig1} (b), for the Flqouet state at $J_0(A/\omega )=0$, we show the probability of finding the particle in different lattice sites. Generally, the Floquet states are either localized in the site 1 or equally distributed on both site 2 and site 3, which confirm the approximate Floquet solutions very well.

\begin{figure}[htp]
\center
\includegraphics[width=1.0\columnwidth]{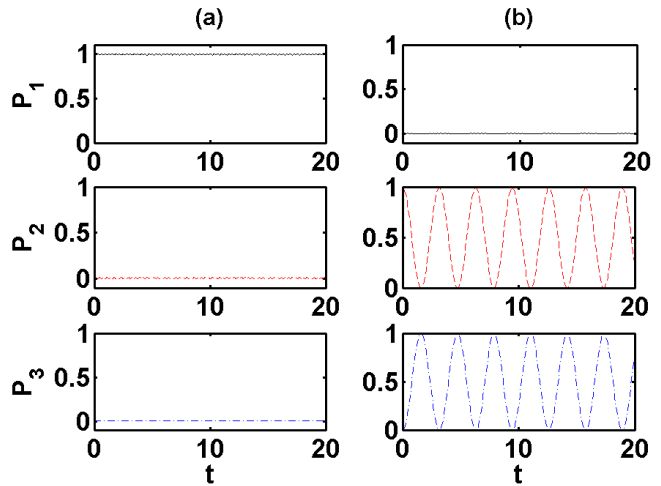}
\caption{(color online) Time evolution of the probability distribution for the system of $A/\omega\simeq 2.405$. The particle is initialized in site 1 [column (a)], and site 2 [column (b)]. The modulations and the parameters are the same as the ones for Fig. \ref{fig1}.} \label{fig2}
\end{figure}

Our numerical simulation is implemented by directly integrating the time-dependent
Scr\"{o}dinger equation (\ref{eq:c3}) with type-(a) or type-(b) modulations. For different initial conditions, we calculate the probabilities $P_j=|a_j|^2, (j=1,2,3)$ of finding the particle in lattice site $j$. In Fig. \ref{fig2}(a), we show the time evolution for the system of $A/\omega\simeq 2.405$ and the particle is initially localized in the site 1. Since the tunnelling between the site 1 and the other two sites are suppressed, the particle is frozen in the site 1. On the other hand, if we start a particle in the site 2, the particle indeed simply tunnels back and forth between the sites 2 and 3, see Fig. \ref{fig2} (b). These results indicate that only the tunneling between the site 2 and 3 is allowed. This is a kind of selective CDT, which is also evidenced by the probability distribution of the Flqouet states shown in Fig.~\ref{fig1}~(b). One can use the effective  Scr\"{o}dinger equation (\ref{eq:ceff}) to explain the SCDT. Under the condition of $J_0(A/\omega )=0$ and $b_1=1, b_2=b_3=0$, one can immediately give the analytical solution as $b_1=1,b_2=b_3=0$. Similarly, if the initial condition takes $b_1=0, b_2=1,b_3=0$, the analytical solution is given as $b_1=0$, $b_2=\cos(\Omega t)$ and $b_3=-i\sin(\Omega t)$.

\begin{figure}[htp]
\center
\includegraphics[width=1.0\columnwidth]{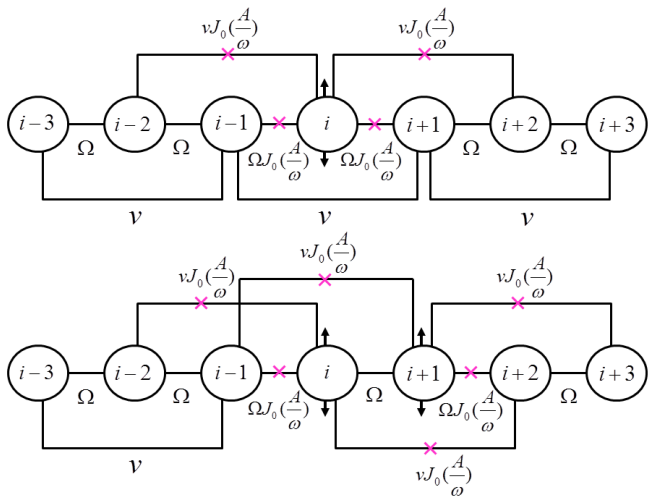}
\caption{(color online) The schematic diagram for the dynamics of lattice arrays under selective in-phase harmonic modulations. Upper cartoon: the modulation is only applied to site $i$ (marked by $\updownarrow$), that is, $\varepsilon_i(t)=A\sin(\omega t)$ and $\varepsilon_j(t)=0$ for $(j\neq i)$. Lower cartoon: in-phase modulations are applied to sites $i$ and $i+1$ (marked by $\updownarrow$), that is, $\varepsilon_i(t)=\varepsilon_{i+1}(t)=A\sin(\omega t)$ and $\varepsilon_j(t)=0$ for $(j\neq i,i+1)$.} \label{fig3}
\end{figure}

Although the above analysis is applied to three-site systems, similar analysis can be applied to systems of arbitrary number of lattice sites and similar CDT phenomena can be found. The effective inter-site tunnelings are illustrated by the cartoon shown in Fig.~\ref{fig3}. If system is driven by a high-frequency field such that $J_0(A/\omega )=0$, as illustrated in Fig.~\ref{fig3}, then the coupling parameters between the unmodulated and modulated site are effectively rescaled by the factor $J_0$, while the couplings among modulated sites or unmodulated sites keep unchanged. The upper cartoon in Fig.~\ref{fig3} represents the systems of modulation on site $i$ only, in which the site $i$ becomes a true trap for the particle, since tunneling in both directions are suppressed, which is marked by the red cross. If in-phase modulations are applied to both sites $i$ and $i+1$, the lattice array divides into three sets of disconnect sites $(...,i-3,i-2,i-1)$, $(i,i+1)$ and $(i+2,i+3,...)$, see the lower cartoon. If the particle is initialized in site $i$ or $i+1$, Rabi oscillation will occur between site $i$ and $i+1$, and the frequency of this oscillation is determined by the value of the original tunneling strength $\Omega$.

\section{Possible applications}

\begin{figure}[htp]
\center
\includegraphics[width=1.0\columnwidth]{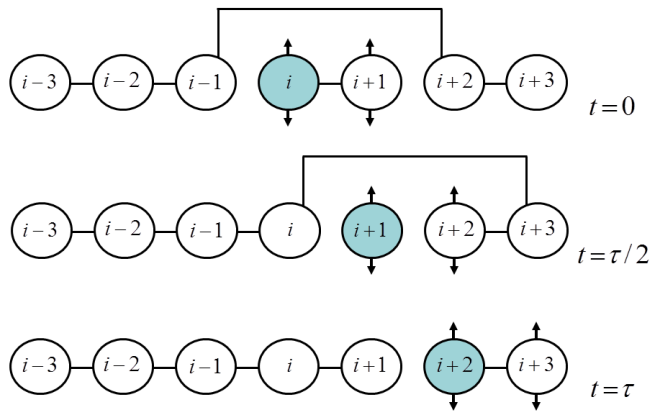}
\caption{(color online) Schematic diagram for transporting a
particle from left to right in a lattice array. The symbol
$\updownarrow$ denotes the high-frequency modulation
$\varepsilon=A\sin(\omega t)$, in which $A/\omega \simeq 2.4$. After
a time interval of $t=\tau/2$ with $\tau=\pi/\Omega$, the particle
has completely transported from one site to its right-side
neighbor.} \label{fig4}
\end{figure}

Based on the CDT in a lattice system under selective in-phase modulations, we propose a scheme for directed single-particle transport, which can be used to realize a quantum motor. The transport process is illustrated in Fig.~\ref{fig4}, in which $A/\omega\simeq 2.405$ and $\omega/\Omega=10$. The particle is initially loaded into site $i$ and in-phase
modulations are applied to sites $i$ and $i+1$, see Fig.~\ref{fig4}. The Rabi oscillation between sites $i$ and $i+1$ has a period $\tau=\pi/\Omega$. At $t=\tau/2$, the particle arrives at site $i+1$ and the in-phase modulations are suddenly changed to address sites $i+1$ and $i+2$. Therefore the particle enters into a new Rabi oscillation between sites $i+1$ and $i+2$ with the same period $\tau$. Similarly, at $t=\tau$, the modulations are suddenly switched to address sites $i+2$ and $i+3$, the particle has completely tunneled to site
$i+3$ after another time interval of $\tau/2$. If the in-phase modulations are repeatedly applied to the lattice site occupied by the particle and its right-side neighbor at
$t=n\tau/2$ for $n=0,1,2,3...$, the particle will propagate solely to the right. Similarly, the in-phase modulations are repeatedly applied to the lattice site occupied by the particle and its left-side neighbor at $t=n\tau/2$ for $n=0,1,2,3...$, the particle will propagate solely to the left.

\begin{figure}[htp]
\center
\includegraphics[width=1.0\columnwidth]{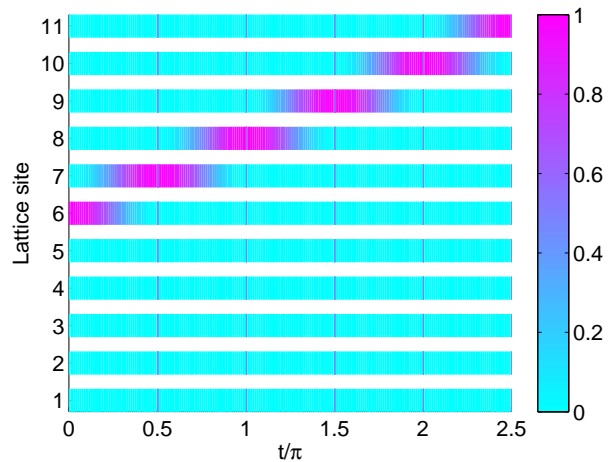}
\caption{(color online) Simulation of a single particle in a 11-site
system. The situation is illustrated in Fig \ref{fig4}. }
\label{fig5}
\end{figure}

The directed quantum transport based upon CDT provides an optional scheme for designing a quantum motor. To illustrate the possibility of realizing a quantum motor via the directed quantum transport, we show our numerical results of a single particle in a 11-site system with the procedure described above, see Fig.~\ref{fig5}. Obviously, the directed single-particle transport is perfectly confirmed by our results.

In addition to the application in designing a quantum motor, the CDT may be used to realize a quantum beam splitter. If the in-phase modulations are initially applied to the sites $i-1$, $i$ and $i+1$, that is, $\varepsilon_{i-1}=\varepsilon_{i}=\varepsilon_{i+1}=A\sin(\omega
 t)$ and $\varepsilon_j(t)=0$ for $j\neq i-1, i, i+1$. Therefore the three sites
$i-1$, $i$ and $i+1$ are disconnected with other sites, and the couplings
among the three sites keeps unchanged. The dynamics can be well
effectively described by Eq.~(\ref{eq:c3}) for the case of no modulations  (i.e., $\varepsilon_1=\varepsilon_2=\varepsilon_3=0$). Initializing the system into a state of
$a_2=1$ and $a_1=a_3=0$, we can derive the analytical solution for Eq.~(\ref{eq:c3}) with
$\varepsilon_1=\varepsilon_2=\varepsilon_3=0$, which has equal probability amplitudes for the first and third sites($P_1=P_3$). The particle has same probability for tunneling to sites $i-1$ and $i+1$ and therefore the state splits into two parts. At the time of zero probability for finding the particle in the site $i$, we switch off the in-phase modulations on the site $i$ and simultaneously switch on in-phase modulations on sites $i-2$ and $i+2$. After a time interval of $\tau/2$, one will find the particle in site $i-2$ or  $i+2$ with equal probability. After a time interval of $\tau/2$, we switch off the modulations on sites $i-2$ and $i+2$ and simultaneously switch on the modulations on sites $i-3$ and $i+3$. Again and again, after each time interval of $\tau/2$, we switch off the modulations on sites $i-k$ and $i+k$ and simultaneously switch off the modulations on sites $i-(k+1)$ and $i+(k+1)$. Under this sequence of applying the in-phase modulations to some certain lattice sites, the particle initially localized in the central site will form an equal-probability superposition of left and right propagating components. This means that a perfect quantum beam splitter is realized by the CDT caused by the in-phase modulations.

\begin{figure}[htp]
\center
\includegraphics[width=1.0\columnwidth]{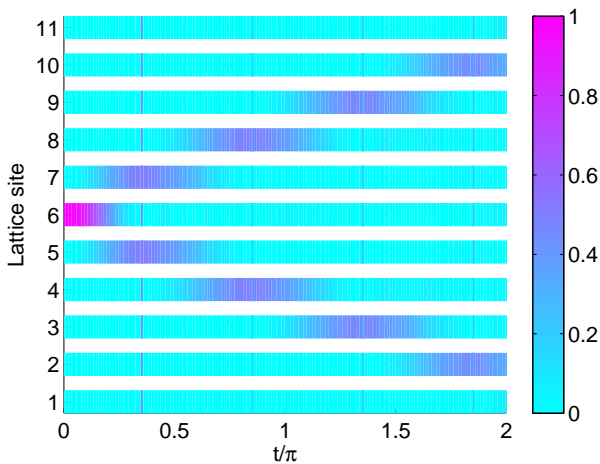}
\caption{(color online) A quantum beam splitter for a particle in a 11-site system. If the in-phase modulations initially acts on the site 6 and its nearest neighbors (the sites 5 and 7), the state splits into two equal parts propagating along opposite directions.}
\label{fig6}
\end{figure}

Although some other good schemes for directed transport of single-particle has been previously proposed in a driven lattice~\cite{Creffield}, we stress here that one advantage of our approach is that it can be applied to lattices with high-order coupling among non-nearest sites.

\section{Optical realization}

By mapping the temporal evolution of quantum systems into the spatial propagations of light waves, the engineered waveguides offer an alternative platform to investigate the classical
wave analogues of a wide variety of coherent quantum effects. Below, we discuss how to simulate the CDT induced by selective in-phase modulations via the engineered waveguides.

The propagation of a cw light wave along the $z$ axis is described as the Schr\"{o}dinger equation for its dimensionless field amplitude $\phi(x,z)$\cite{Szameit,Kartashov}
\begin{eqnarray}
i\frac{\partial\phi}{\partial
z}=-\frac{1}{2}\frac{\partial^2\phi}{\partial x^2}-pR(x,z)\phi.
 \label{eq:schopt}
\end{eqnarray}
Here ($x$ and $z$) are the normalized (transverse and longitudinal) coordinates, and $p$ describes the refractive index contrast of the individual waveguide. The refractive index distribution of the waveguide lattice is given by
\begin{eqnarray}
R(x,z)&=&\sum_{j=-M}^{M}[1+f_{j}(z)]\exp\left[-\left(\frac{x-jw_s}{w_x}\right)^6\right],\\
f_{j}(z)&=&F_j\mu\sin\left(\omega z\right),\nonumber
\end{eqnarray}
with the waveguide spacing $w_s$, the channel width $w_x$, the longitudinal modulation amplitude $\mu$, and the modulation frequency $\omega$. The super-Gaussian function $\exp(-x^6/w_x^6)$ describes the profile of individual waveguides with widths $w_x$.  $2M+1$ is the total number of the waveguides. The function $F_j$ determines the modulation type. Here we consider the selective in-phase harmonic modulations, such that the value of $F_j$ can only be either $F_j=1$ or $F_j=0$. The case of $F_j=1$ indicates that the $j$-th waveguide is modulated, while $F_j=0$ means that the $j$-th waveguide is unmodulated.

Under the tight-binding condition, one can expand the total electric envelope in form of
\begin{eqnarray}
\phi(x,z)&=&\sum_j a_j\phi_j(x)\exp(-i\beta_0 z),\\
\beta_0&=&\int \phi_j^*(x)H_0 \phi_j(x)dx,\nonumber\\
H_0&=&-\frac{1}{2} \frac{\partial^2}{\partial x^2} -p\sum_{j=-M}^{M}\exp\left[-\left(\frac{x-jw_s}{w_x}\right)^6\right],\nonumber
\end{eqnarray}
where $\phi_j(x)$ are the eigen-modes for the $j$-th waveguide, and the expanding coefficients are normalized to one, $\sum_j|a_j|^2=1$. Substituting the expansion (10) into the Schr\"{o}dinger equation~(\ref{eq:schopt}) and taking into account couplings between next-nearest and nearest waveguides, one can obtain the resulting coupled-mode equation in the same form as Eq.~(\ref{eq:c}) with $t$ replaced by $z$. Therefore, the temporal evolution of the probability amplitudes $a_j$ obeying Eq. (\ref{eq:c}) can be simulated by the light propagation in an array of tight-binding waveguides with longitudinal modulation, in which the temporal evolution of $a_j$ is mapped into the spatial evolution of the modal field amplitudes of light waves in the $j$-th waveguide along the axial direction $x$. The light power distribution $P(x,z)$ corresponds to the probability distribution $|a_j(t)|^2$.

To demonstrate the quantum control based on the CDT, we simulate the three-channel couplers ($M=1$) by integrating the field propagation equation~(\ref{eq:schopt}). In our numerical simulation, the initial states are chosen as the lowest Wannier modes for isolated individual waveguides, and the parameters are set as $w_x=0.3$, $w_s=3.2$ and $p=2.78$. As same as the units used in Refs.~\cite{Szameit,Kartashov}, $w_x$ and $w_s$ are in units of 10 $\mu\textrm{m}$, and $p=2.78$ corresponds to a real refractive index of $3.1 \times 10^{-4}$.

\begin{figure}[htp]
\center
\includegraphics[width=1.0\columnwidth]{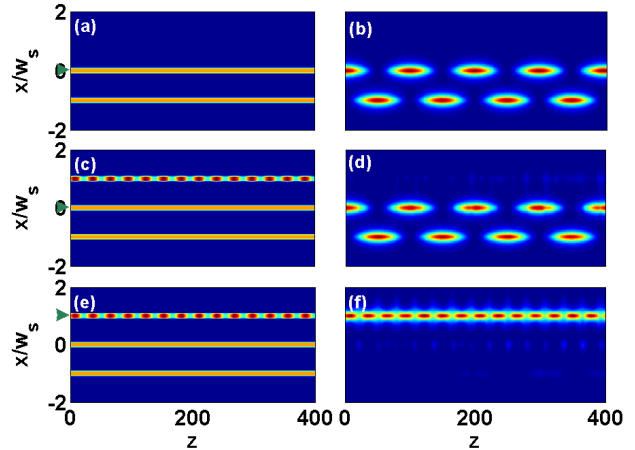}
\caption{(color online). Light propagation in a three-channel coupler under type-(i) modulation. First row:  (a) the refractive index distribution $R(x,z)$ and (b) the light propagation $|\phi(x,z)|^2$ for the unmodulated two-waveguide system corresponding to the two unmodulated waveguides decoupled from the original three-channel system. Second row: (c) $R(x,z)$ and (d) $|\phi(x,z)|^2$ with the input beam centered in the middle waveguide marked by $\blacktriangleright$ in (c). Third row: (e) $R(x,z)$ and (f) $|\phi(x,z)|^2$ with the input beam centered in the top waveguide marked by $\blacktriangleright$ in (e).} \label{fig7}
\end{figure}

According to the analytical analysis in Section III, if not all three waveguides are modulated simultaneously and modulations satisfy $J_0(A/\omega)=0$, the three-channel couplers can be decoupled into a coupled two-waveguide system and an isolated single-waveguide system. To compare with the sub-systems decoupled from the original system, we also simulate the tunneling dynamics of light in an unmodulated two-waveguide system corresponding to the two waveguides decoupled from the original system, see Figs.~\ref{fig7}(b).

Similar to the discrete systems, the in-phase modulations applied to the continuous system (\ref{eq:c}) may suppress the tunneling between modulated and unmodulated waveguides for some specific modulation amplitudes and frequencies. We find that the tunneling between modulated and unmodulated waveguides is strongly suppressed if the modulation frequency $\omega=3.45 \times (2\pi/100)$ and $\mu=0.4$. This strong suppression of the tunneling between modulated and unmodulated waveguides is a new extension of CDT.

\begin{figure}[htp]
\center
\includegraphics[width=1.0\columnwidth]{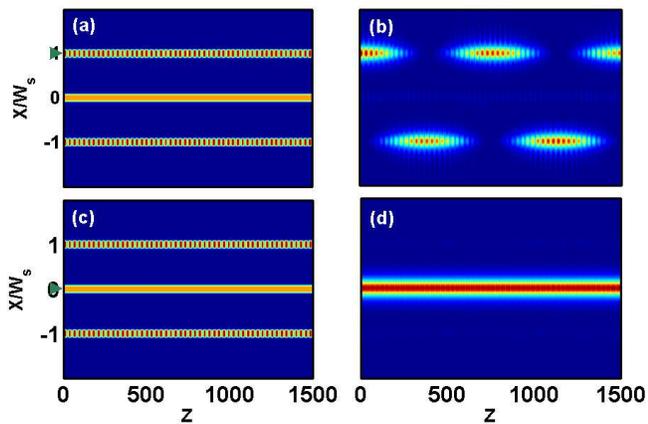}
\caption{(color online) Light propagation in a three-channel coupler under type-(ii) modulation. First row: (a) $R(x,z)$ and (b) $|\phi(x,z)|^2$ with the input beam centered in the top waveguide marked by $\blacktriangleright$ in (a). Second row: (c) $R(x,z)$ and (d) $|\phi(x,z)|^2$ with the input beam centered in the middle waveguide marked by $\blacktriangleright$ in (c).} \label{fig8}
\end{figure}

We simulate a three-channel coupler under two typical in-phase modulations: (i) in-phase modulation is only applied to one waveguide [see Figs.~\ref{fig7} (c) and (e)], i.e. $F_1=1$ and $F_{-1}=F_0=0$; and (ii) in-phase modulations are applied to two next-nearest-neighboring waveguides [see Figs.~\ref{fig8} (a) and (c)], i.e. $F_1= F_{-1}=1$ and $F_0=0$. We numerically explore the light propagation in a three-channel coupler under these two modulations. In these two figures, the left columns show the refractive index distributions $R(x,z)$ and the right columns show the light propagation $|\phi(x,z)|^2$ which are obtained by numerically integrating the Schr\"{o}dinger equation~(\ref{eq:schopt}). In our simulation, the parameters are set as $w_x=0.3$, $w_s=3.2$, $p=2.78$, $\mu=0.4$ and $\omega=3.45\times(2\pi/100)$.

In Fig. 7 (c)-(f), for a three-channel coupler under type-(i) modulation, we show the refractive index distribution $R(x,z)$ [see panels (c) and (e)] and the light propagation $|\phi(x,z)|^2$ [see panels (d) and (f)] for different input beams marked by $\blacktriangleright$. The first row shows $R(x,z)=\exp[-x^6/w_x^6]+\exp[-(x+w_s)^6/w_x^6]$ and $|\phi(x,z)|^2$ for the unmodulated two-waveguide system corresponding to the subsystem decoupled from the original three-channel system, see Figs. 7 (a) and (b). If the input beam is centered in the middle waveguide marked by $\blacktriangleright$ [see panel (c)], which is one of two unmodulated waveguides, the light periodically tunnels between the two unmodulated waveguides. The tunneling dynamics is almost the same as the one in the unmodulated two-waveguide coupler with $R(x,z)=\exp[-x^6/w_x^6]+\exp[-(x+w_s)^6/w_x^6]$, see Fig.~\ref{fig7}(b). While the input beam is centered in the sole modulated waveguide marked by $\blacktriangleright$ in Fig.~\ref{fig7} (e), the light propagates in this modulated waveguide almost without tunneling into the two unmodulated waveguides, see Fig.~\ref{fig7} (f). The light propagation shown in Figs.~\ref{fig7} (d) and (f) clearly shows the strong suppression of the coupling between modulated and unmodulated waveguides.

For a three-channel coupler under type-(ii) modulation, in which the top and bottom waveguides are modulated, we show the corresponding light propagation for different input beams in Figs.~\ref{fig8} (a) and (c). Similarly, if the input beam is centered in one of the two modulated waveguides, the light only oscillates between the two modulated waveguides, see Fig.~\ref{fig8} (a) and (b). While the input beam is centered in the middle waveguide marked by $\blacktriangleright$ in Fig.~\ref{fig8} (c), the light almost perfectly keeps propagating in this waveguide, see Fig.~\ref{fig8} (d). Obviously, the numerical results indicate the decoupling between the modulated and unmodulated waveguides.

For a quantum lattice system, it is generally thought that the nearest-neighboring tunneling is more significant than the next-nearest-neighboring one. However, due to the in-phase periodic modulations, even the nearest-neighboring tunneling strength is stronger than the next-nearest-neighboring one, the appearance of significant tunneling between next-nearest-neighboring waveguides and almost no tunneling between nearest-neighboring waveguides updates this conventional understanding. This provides an optional approach for directly transporting a particle to its non-nearest-neighboring sites.

For different input beams, we have also numerically explored the light propagation $\left|\phi(x, z)\right|^2$ in a three-channel coupler under in-phase modulations applied to two nearest-neighboring waveguides. Again, the light propagation (not shown here) clearly indicates that the modulated and unmodulated waveguides are almost decoupled each other and there is almost no tunneling between them.

Moreover, we have simulated the multi-waveguide systems of other numbers of waveguides by directly integrating the Schr\"{o}dinger equation~(\ref{eq:schopt}). The predictions from the coupled-mode equation (\ref{eq:c}) are demonstrated and confirmed perfectly in this optical system (not shown here). Therefore, such a waveguide array under in-phase modulations may be used as a switch-like device, which can function as input-dependent single-mode fibers, dual-channel couplers and multi-channel couplers as well.

\section{Summary}

In conclusion, we have studied how to control the tunneling dynamics in one-dimensional quantum lattice systems via CDT induced by selective in-phase modulations, in which only some selected lattice sites but not all lattice sites are driven by periodic in-phase fields. Tuning the frequency and amplitude of the selective in-phase modulations to some specific values, it is possible to find that the CDT only occurs between modulated and unmodulated sites while the tunneling dynamics among modulated or unmodulated sites keeps unchanged. In particular, by applying in-phase modulations to next-nearest-neighboring sites, it is possible to switch off the tunneling between the nearest-neighboring sites. Therefore, one can utilize the in-phase modulations to produce ratchet-like motions and even to transport directly a particle between non-nearest-neighboring lattice sites with the high-order coupling among those lattice sites.

Furthermore, we have shown that our results on CDT can be experimentally tested with an array of optical waveguides, in which the spatial propagation of light in an optical material system corresponds to the temporal evolution of state in a quantum system. Comparing with previous works on light propagation in optical waveguides under out-phase modulations~\cite{Szameit,Kartashov}, the in-phase modulations provide a new possibility for controlling light propagation in periodic waveguide arrays.

\section*{Acknowledgments}
The authors thank B. Wu for his helpful discussions. This work is supported by the NNSFC under Grants No. 10965001, 11165009 and 11075223,  the Natural Science Foundation of Jiangxi Province under Grant No. 2010GQW0033, the Jiangxi young scientists training plan under Grant No. 20112BCB23024, the NBRPC under Grant No. 2012CB821300 (2012CB821305), the NCETPC under Grant No. NCET-10-0850 and the Fundamental Research Funds for Central Universities of China.


\begin{thebibliography}{99}
\bibitem{Grifoni} M. Grifoni et al., Phys. Rep. \textbf{304}, 229(1998).

\bibitem{Kohler} S. Kohler et al., Phys. Rep. \textbf{406}, 379(2005).

\bibitem{Grossmann} F. Grossmann, T. Dittrich, P. Jung, and P. H\"{a}nggi, Phys. Rev. Lett. \textbf{67}, 516 (1991); F. Grossmann, P. Jung, T. Dittrich, and P. H\"{a}nggi, Z. Phys. B \textbf{84}, 315 (1991).

\bibitem{Dunlap} D. H. Dunlap and V. M. Kenkre, Phys. Rev. B \textbf{34}, 3625 (1986).

\bibitem{Villas} J. M. Villas-B\^{o}as, S. E. Ulloa, and N. Studart, Phys. Rev. B \textbf{70}, 041302(R) (2004).

\bibitem{Stockburger} J. T. Stockburger, Phys. Rev. E \textbf{59}, R4709 (1999).

\bibitem{Luo} X. Luo, Q. Xie, B. Wu,  Phys. Rev. A \textbf{76}, 051802(R)(2007); X. Luo, Q. Xie, B. Wu, Phys. Rev. A \textbf{77}, 053601(2008).

\bibitem{Gong} J. Gong, L. Morales-Molina, and P. H\"{a}nggi, Phys. Rev. Lett. \textbf{103}, 133002 (2009).

\bibitem{Kierig} E. Kierig, U. Schnorrberger, A. Schietinger, J. Tomkovic, and M. K. Oberthaler, Phys. Rev. Lett. \textbf{100}, 190405 (2008).

\bibitem{Lignier} H. Lignier, C. Sias, D. Ciampini, Y. Singh, A. Zenesini, O. Morsch, and E. Arimondo, Phys. Rev. Lett. \textbf{99}, 220403 (2007).

\bibitem{Salger} T. Salger, S. Kling, T. King, C. Geckeler, L. M. Molina, and M. Weitz, Science \textbf{326}, 1241 (2009).

\bibitem{Hai} G. Lu, W. Hai, Phys. Rev. A \textbf{83}, 053424 (2011).

\bibitem{Romero} O. Romero Isart and J. J.Garc\'{\i}a-Ripoll, Phys. Rev. A \textbf{76}, 052304 (2007).

\bibitem{Creffield} C. E.Creffield, Phys. Rev. Lett. \textbf{99}, 110501 (2007).

\bibitem{Longhi} S. Longhi, Laser and Photon. Rev. \textbf{3}, 243 (2008).

\bibitem{Longhi2} S. Longhi, M. Marangoni, M. Lobino, R. Ramponi, P. Laporta, E. Cianci, and V. Foglietti, Phys. Rev. Lett. \textbf{96}, 243901 (2006).

\bibitem{Valle} G. Della Valle, M. Ornigotti, E. Cianci, V. Foglietti, P. Laporta and S. Longhi, Phys. Rev. Lett. \textbf{98}, 263601 (2007).

\bibitem{Szameit}A. Szameit, Y. V. Kartashov, F. Dreisow, M. Heinrich, T. Pertsch S. Nolte, A. T\"{u}nnermann, V. A. Vysloukh, F. Lederer, and L. Torner, Phys. Rev. Lett. \textbf{102}, 153901 (2009); A. Szameit Y. V. Kartashov, M. Heinrich, F. Dreisow, R. Keil, S. Nolte A. T\"{u}nnermann, V. A. Vysloukh, F. Lederer, and L. Torner, Opt Lett. \textbf{34}, 2700 (2009).

\bibitem{Gong-PRA-2007} J. B. Gong, D. Poletti, and P. Hanggi, Phys. Rev. A \textbf{75}, 033602 (2007).

\bibitem{Kartashov} Y. V. Kartashov and V. A. Vysloukh, Opt. Lett. \textbf{34}, 3544 (2009); Y. V. Kartashov, A. Szameit, V. A. Vysloukh, and L. Torner, Opt. Lett. \textbf{34}, 2906 (2009); Y. V. Kartashov and V. A. Vysloukh, Opt. Lett. \textbf{35}, 2097(2010).

\end{thebibliography}
\end{document}